\documentclass[preprint,12pt]{elsarticle}

\usepackage{amssymb}
\usepackage{multirow}
\usepackage{booktabs}
\usepackage{amssymb}
\usepackage{mathrsfs}
\usepackage{longtable}
\usepackage{lscape}
\usepackage{tabularx}
\usepackage{epsfig}
\usepackage{colortbl}
\usepackage{subfigure}
\usepackage[table]{xcolor}
\usepackage{float}
\journal{The Scientific WorldJournal}

\begin{document}

\begin{frontmatter}



\title{A betweenness structure entropy of complex networks}


\author[swu]{Qi Zhang}
\author[swu]{Meizhu Li}
\author[swu,vu,NWPU]{Yong Deng\corref{cor}}
\ead{ydeng@swu.edu.cn, prof.deng@hotmail.com}

\cortext[cor]{Corresponding author: Yong Deng, School of Computer and Information Science, Southwest University, Chongqing, 400715, China.}

\address[swu]{School of Computer and Information Science, Southwest University, Chongqing, 400715, China}
\address[NWPU]{School of Automation, Northwestern Polytechnical University, Xian, Shaanxi 710072, China}
\address[vu]{School of Engineering, Vanderbilt University, Nashville, TN, 37235, USA}

\begin{abstract}
The structure entropy is an important index to illuminate the structure property of the complex network. Most of the existing structure entropies are based on the degree distribution of the complex network. But the structure entropy based on the degree can not illustrate the structure property of the weighted networks. In order to study the structure property of the weighted networks, a new structure entropy of the complex networks based on the betweenness is proposed in this paper.
Comparing with the existing structure entropy, the proposed method is more reasonable to describe the structure property of the complex weighted networks.

\end{abstract}
\begin{keyword}
complex networks \sep structure entropy \sep betweenness \sep weighted network


\end{keyword}
\end{frontmatter}

\section{Introduction}
The complex networks is a graph with non-trivial topological features, the feature that do not occur in simple networks but often occur in real networks. Many real networks are the complex networks, such as the social networks, information networks, technological networks and biological networks \cite{newman2003structure}. Recently, many researcher have been interested to explore the complex networks. In 1998, Watts and Strogatz proposed the principle of 'Small-world' for the complex networks on Nature \cite{watts1998collective}. Then the 'Scale-free networks' is proposed by some researchers \cite{barabasi2000scale}. Then the statistical theory is introduced in the complex networks \cite{berg2002correlated,albert2002statistical}. Those researches have revealed that the structure property is important to research the complex networks.




Many of the existing structure entropies are based on the degree distribution of the complex networks. But the degree of the complex networks is a $local$ measure to some degree, ignoring the influence of the edge's weighted in the structure property. As a results, the structure entropy based on the degree can not describe the structure property of those complex weighted networks, especially for those networks with a uniform degree distribution and different weighted of the edges. To describe the structure property of those complex weighted networks, we need to find a new method to represent the structure entropy.

Compared with the degree measure of the complex networks, the betweenness is a global measure of complex networks. It is defined based on the shortest path of the networks. It can be used to describe the structure property of the complex networks from the global view.

In this paper, we proposed a new structure entropy of the complex networks which is based on the betweenness of the complex networks and the information theory. The results of our research have revealed that the structure entropy based on the betweenness is a useful method to describe the structure property of the complex networks.

The rest of this paper is organised as follows. Section \ref{Rreparatorywork} introduces some preliminaries of this work. In section \ref{new}, a new structure entropy of the complex networks based on the betweenness is proposed. The application of the proposed method is illustrated in section \ref{application}. Conclusion is given in Section \ref{conclusion}.
\section{Preliminaries}
\label{Rreparatorywork}
\subsection{Betweenness}
\label{betweennessdefine}
The betweenness is an important index which can be used to illuminate the importance of the nodes. It is defined based on the shortest path of the network \cite{barthelemy2004betweenness}.

\begin{figure}[H]
  \centering
  \includegraphics[scale=0.1]{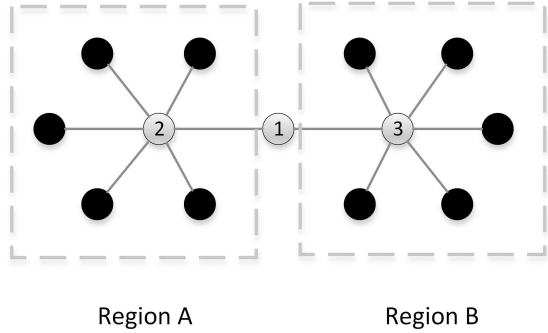}\\
  \caption{The vertex 1 in the figure connect the region A and region B. The degree of the vertex 1 is 2, but all shortest path from the nodes in region A to the nodes in region B have to go to through the vertex 1. The vertex 1 have a small value of degree but a large value of betweenness. In fact, vertex 1 is here a cut-vertex; its removal will break the network into two disconnected components.}\label{betweenness}
\end{figure}

The betweenness of the complex networks is defined as follows \cite{barthelemy2004betweenness}:

\begin{equation}\label{Betweenness}
bet(i) = \frac{{\upsilon (i)}}{{\sum {{\sigma _{st}}} }}(s \ne i \ne t)
\end{equation}

In the Eq. (\ref{Betweenness}), the ${{\sigma _{st}}}$ is the number of the shortest path from vertex $s$ to vertex $t$, ${\upsilon (i)}$ is the number of the shortest path which have go to through the vertex $i$ \cite{barthelemy2004betweenness}.

\subsection{Existing structure entropy}
\label{entropy}
The structure entropy of the complex networks is based on the information entropy \cite{shannon2001mathematical} and the statistic characteristics of the complex networks. It can be used to describe the structure property of the complex networks.

The information entropy is a conception of information theory which is proposed by Shannon \cite{shannon2001mathematical}. Shannon defined the information as "the reduction of entropy", "the reduction of uncertainty of a system", and firstly proposed the quantitative description method for information.

Suppose $X = \{ {x_1},{x_2},{x_3}, \cdots ,{x_n}\}$ is a discrete random variable, the appearance probability of information source given by $X$ is denoted as
$p_i  = p(x_i ),i = 1,2, \ldots ,n$, and $\sum\limits_{i = 1}^n {{p_i}}  = 1$. Then the information entropy is defined as follows:

\begin{equation}\label{information-entropy}
H =  - k\sum\limits_{i = 1}^n {{p_i}\log {p_i}}
\end{equation}
Where $k$ is equal to 1, $n$ is the number of the probabilities.

Many researchers have proposed the methods to calculate the structure entropy of the complex networks, such as the structure entropy based on the degree distribution  \cite{wang2006entropy}, the structure entropy based on the automorphism partition of the network  \cite{xiao2008symmetry} and the structure entropy based on the degree dependence matrices \cite{xu2013degree}. Most of those structure entropies are based on the degree of the nodes, defined as follows \cite{wang2006entropy}:

\begin{equation}\label{degree-entropy}
H_{deg} =  - k\sum\limits_{j = 1}^N {{p_j}\log {p_j}}
\end{equation}

Where the ${p_j}$ is defined as follows:
\begin{equation}\label{pj}
{p_j} = {\textstyle{{Degree(j)} \over {\sum\limits_{j = 1}^N {Degree(j)} }}}
\end{equation}
Where the $Degree(j)$ represent the $j$th vertex's degree and $N$ is the total number of the nodes in the network.

The structure entropy based on the automorphism partition of the network is defined as follows \cite{xiao2008symmetry}:

\begin{equation}\label{automorphism-degree-entropy}
H_{partition} =  -\sum\limits_{p = 1}^{\left| P \right|} {{p_p}\log {p_p}}
\end{equation}

Where $P$ is the automorphism partition of the network, ${p_p}$ is the probability that a vertex belongs to the cell ${V_i}$ of the $P$. Note that given a network's automorphism partition $P = \left\{ {{V_1},{V_2},{V_3}, \ldots ,{V_k}} \right\}$, the ${p_p}$ is calculated as:

\begin{equation}\label{pp}
{p_p} = {\textstyle{{\left| {{V_p}} \right|} \over {\sum\limits_{p = 1}^k {\left| {{V_p}} \right|} }}} = {\textstyle{{\left| {{V_p}} \right|} \over N}}
\end{equation}

\begin{figure}
  \centering
  \includegraphics[scale=3.5]{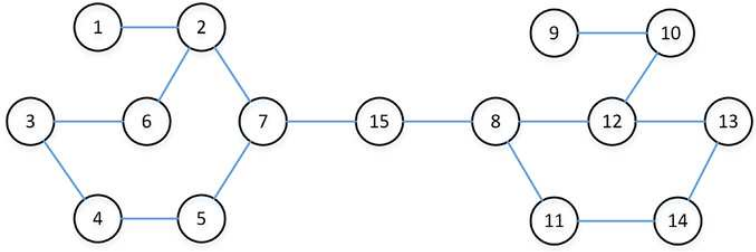}\\
  \caption{In this graph, we can partition the network based on the degree. The degree partition $D = \{ \{ 1,9\} ,\{ 3,4,5,10,11,13,14,15\} ,\{ 2,7,8,12\} \}$ is a coarser automorphism partition of the networks. In the cell \{1,9\} of degree partition, all vertices have degree 1. In this network $P=D$, and $V_1$=\{1,9\}, $V_2$=\{3,4,5,10,11,13,14,15\}, $V_3$=\{2,7,8,12\}. It is clearly that $p_1=2/15$, $p_2=9/15$ $p_3=4/15$. The structure entropy based on the degree partition of this network $H_{partition}= 0.9276.$  }\label{pp-example}
\end{figure}

Where the $k$ is the cell's mounts of the $P$. The Fig. \ref{pp-example} shows a example about how to calculate the structure entropy based on the automorphism partition of the network.

\subsection{The shortcoming of degree-based structure entropy}
\label{shorcoming}
\begin{figure}[H]
  \centering
  \includegraphics[scale=1.2]{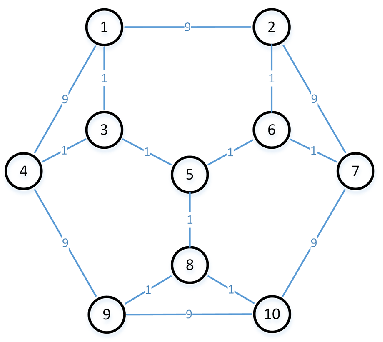}\\
  \caption{The network A}\label{networkB}
\end{figure}

A weighted network is shown in Fig. \ref{networkB}.

The details of the network A is shown in Table \ref{NetworksB-detail}.

\begin{table}[H]
  \centering
  \caption{The betweenness and degree of the network A}
    \begin{tabular}{lccc}
    \toprule

     Node label & degree & betweenness & The number of path across the vertex \\ 
           \midrule
    vertex 1 & 3     & 0.035 & 20 \\   
    vertex 2 & 3     & 0.035 & 20  \\  
    vertex 3 & 3     & 0.3385 & 98  \\  
    vertex 4 & 3     & 0.035 & 20   \\ 
    vertex 5 & 3     & 0.2101 & 65  \\ 
    vertex 6 & 3     & 0.1518 & 50  \\ 
    vertex 7 & 3     & 0.035 & 20   \\ 
    vertex 8 & 3     & 0.1206 & 42  \\  
    vertex 9 & 3     & 0.0195 & 16  \\  
    vertex 10 & 3     & 0.0195 & 16 \\  

    \bottomrule
    \end{tabular}%
  \label{NetworksB-detail}%
\end{table}%

The network A has 10 nodes and 15 edges. Each node's degree is 3, which means that change the value of the edge's weighted, the degree-based structure entropy of the network A is invariable.
\section{Proposed structure entropy}
\label{new}
To address the issue in Fig \ref{networkB}, we proposed a new structure entropy based on the betweenness of the complex networks. It is defined as follows:
\begin{equation}\label{betweenness-entropy}
{H_{bet}} =  - \sum\limits_{i = 1}^n {{p_i}\log {p_i}}
\end{equation}

Where $p_i$ is defined as follows:

\begin{equation}\label{pi}
{p_i} = \frac{{\upsilon (i)}}{{\sum\limits_{i = 1}^n {\upsilon (i)} }}
\end{equation}

Where $\upsilon (i)$ is the betweenness which is defined in section \ref{betweennessdefine}.

To show the necessity of the proposed method, we have calculated the information loss of the network A with the existing structure entropy and the proposed structure entropy. The results are shown in Table \ref{NetworksA-informationloss}.

\begin{table}[H]
  \centering
  \caption{The information loss test for the network A}

\begin{tabular}{lcccccc}
  \toprule
Loss Vertex & $H_{bet}$  & $H_{bet}^{I_{loss}}$      & $H_{deg}$ & $H_{deg}^{I_{loss}}$      & $H_{partition}$      & $H_{partition}^{I_{loss}}$ \\
    \midrule
    The network A & 1.8585 &      & 2.3026 &      & 0     &  \\
     \hline
    Vertex 1 & 1.9641 & -0.1055 & 2.1808 & 0.1218 & 0.6365 & -0.6365 \\
    Vertex 2 & 1.9589 & -0.1004 & 2.1808 & 0.1218 & 0.6365 & -0.6365 \\
    Vertex 3 & 2.0481 & -0.1895 & 2.1808 & 0.1218 & 0.6365 & -0.6365 \\
    Vertex 4 & 1.892 & -0.0335 & 2.1808 & 0.1218 & 0.6365 & -0.6365 \\
    Vertex 5 & 2.1407 & -0.2821 & 2.1808 & 0.1218 & 0.6365 & -0.6365 \\
    Vertex 6 & 1.7638 & 0.0948 &2.1808  & 0.1218 & 0.6365 & -0.6365 \\
    Vertex 7 & 1.7531 & 0.1055 &2.1808  & 0.1218 & 0.6365 & -0.6365 \\
    Vertex 8 & 1.8165 & 0.0420 &2.1808  & 0.1218 & 0.6365 & -0.6365  \\
    Vertex 9 & 1.9774 & -0.1189 & 2.1808 & 0.1218 & 0.6365 & -0.6365 \\
    Vertex 10 & 1.9774 & -0.1189 & 2.1808 & 0.1218 & 0.6365 & -0.6365 \\

  \bottomrule
\end{tabular}
  \label{NetworksA-informationloss}%
\end{table}%

Where the $H_{x}^{I_{loss}}$ represents the information loss of the network A which is calculated with the existing structure entropy and the proposed structure entropy.

The results show that the proposed structure entropy can illuminate the difference of the information loss of the nodes in the network A.

\begin{figure}
  \centering
  \includegraphics[scale=0.6]{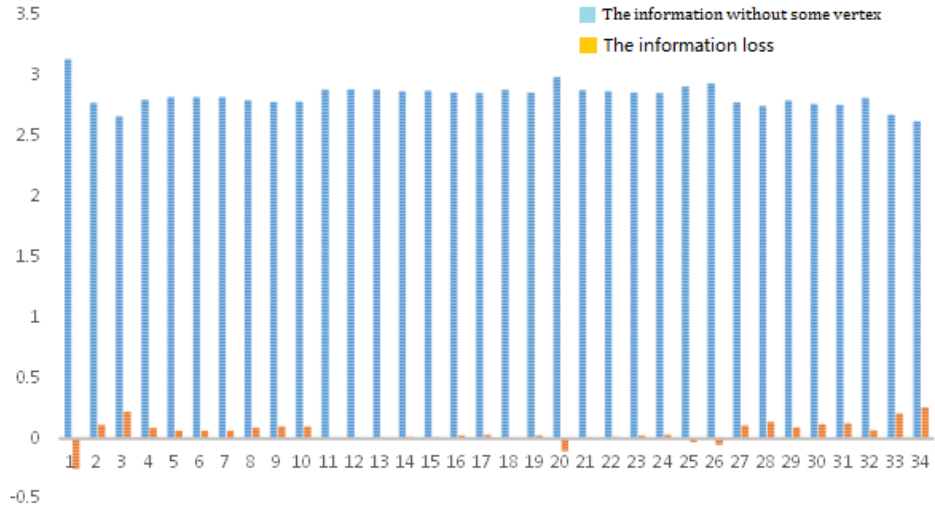}\\
  \caption{The information loss of the Zachary's Karate Club network}\label{information-loss-figure}
\end{figure}

In order to prove the reasonability of the proposed method, the information loss of the Zachary's Karate Club network \cite{uci} is calculated.
The results are shown in Table \ref{Karate-1}, Table \ref{information-loss-karate} and Fig. \ref{information-loss-figure}.

\begin{figure}[htbp]
  \centering
  \includegraphics[scale=0.6]{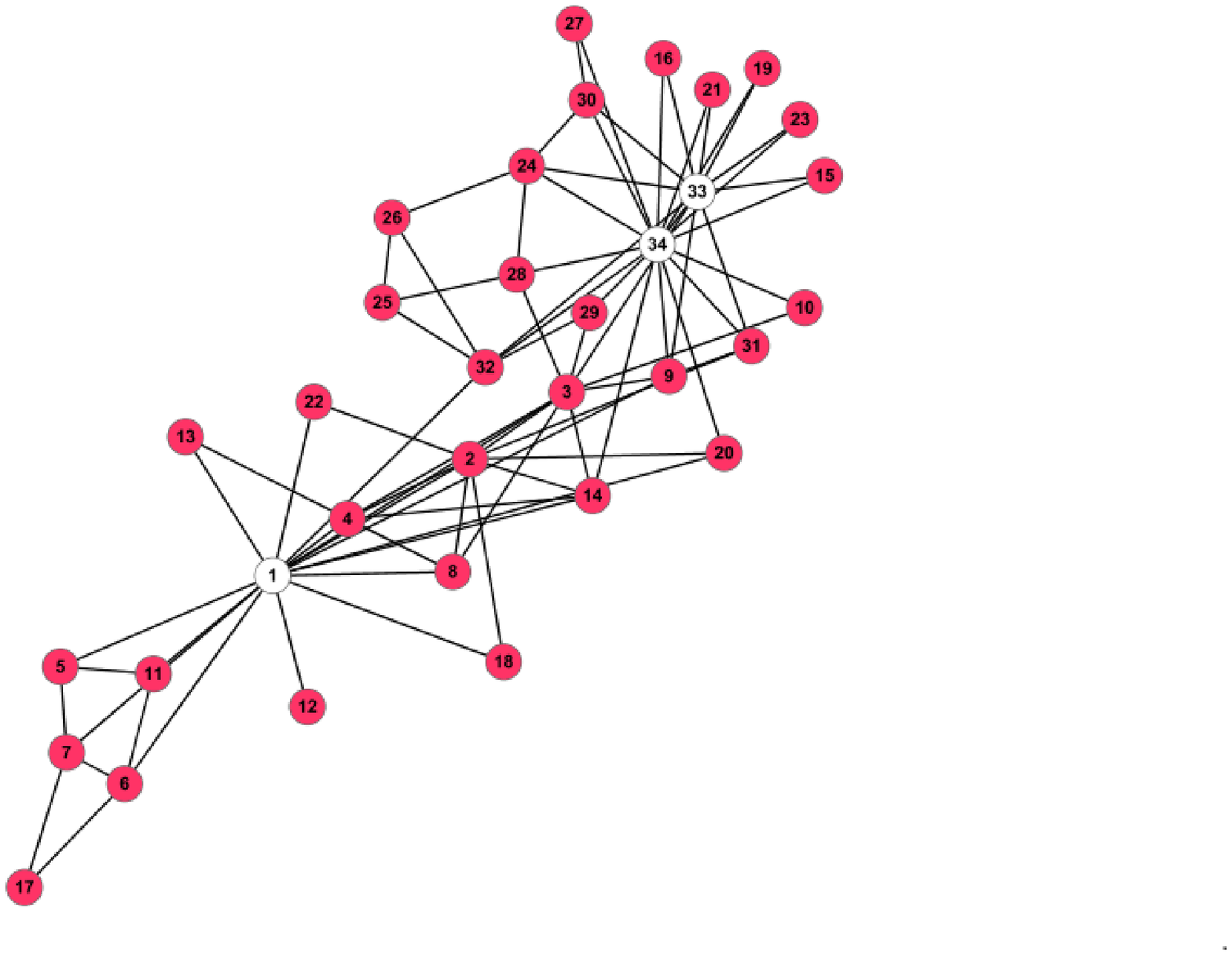}\\
  \caption{The Zachary's Karate Club network}\label{Karate}
\end{figure}

\begin{table}[htbp]
  \centering
  \caption{The details of the Zachary's Karate Club network}
    \begin{tabular}{lcccc}
    \toprule
    Network & Nodes & edges & $C$ & $L$ \\
    \midrule
    Karate & 34    & 78    & 0.4726 & 2.8966 \\
    \bottomrule
    \end{tabular}%
  \label{Karate-1}%
\end{table}%

\begin{table}[H]
\renewcommand{\arraystretch}{0.35}
  \centering
  \caption{The information loss of the Zachary's Karate Club network}
    \begin{tabular}{lccrrrr}
    \toprule
   Loss Vertex &betweenness&degree& $H_{bet}$  & $H_{bet}^{I_{loss}}$  & $H_{deg}$ & $H_{deg}^{I_{loss}}$ \\
    \midrule
     Netwroks         &       &       & 2.8857  &    0   & 3.2609  & 0 \\
    \midrule
    vertex1 & 0.1513  & 16  & 3.1404  & -0.2547  & 3.1970  & 0.0639  \\
    vertex2 & 0.0244  & 9  & 2.7765  & 0.1092  & 3.2031  & 0.0577  \\
    vertex3 & 0.1339  & 10 & 2.6654  & 0.2202  & 3.2198  & 0.0411  \\
    vertex4 & 0.0092  & 6  & 2.8019  & 0.0838  & 3.2194  & 0.0415  \\
    vertex5 & 0.0092  & 3  & 2.8246  & 0.0610  & 3.2247  & 0.0361  \\
    vertex6 & 0.0266  & 4  & 2.8243  & 0.0613  & 3.2117  & 0.0491  \\
    vertex7 & 0.0098  & 4  & 2.8243  & 0.0613  & 3.2117  & 0.0491  \\
    vertex8 & 0.0092  & 4  & 2.8001  & 0.0856  & 3.2357  & 0.0252  \\
    vertex9 & 0.0174  & 5  & 2.7861  & 0.0996  & 3.2401  & 0.0207  \\
    vertex10 & 0.1157  & 2  & 2.7884  & 0.0972  & 3.2433  & 0.0175  \\
    vertex11 & 0.0092  & 3  & 2.8850  & 0.0007  & 3.2247  & 0.0361  \\
    vertex12 & 0.0092  & 1  & 2.8867  & -0.0010  & 3.2490  & 0.0118  \\
    vertex13 & 0.0092  & 2  & 2.8861  & -0.0005  & 3.2393  & 0.0215  \\
    vertex14 & 0.0092  & 5  & 2.8755  & 0.0101  & 3.2411  & 0.0197  \\
    vertex15 & 0.0120  & 2  & 2.8769  & 0.0087  & 3.2446  & 0.0163  \\
    vertex16 & 0.0098  & 2  & 2.8633  & 0.0224  & 3.2446  & 0.0163  \\
    vertex17 & 0.0098  & 2  & 2.8588  & 0.0269  & 3.2265  & 0.0343  \\
    vertex18 & 0.0126  & 2  & 2.8843  & 0.0014  & 3.2422  & 0.0187  \\
    vertex19 & 0.0104  & 2  & 2.8624  & 0.0232  & 3.2446  & 0.0163  \\
    vertex20 & 0.0868  & 3  & 2.9885  & -0.1029  & 3.2433  & 0.0176  \\
    vertex21 & 0.0092  & 2  & 2.8815  & 0.0042  & 3.2446  & 0.0163  \\
    vertex22 & 0.0092  & 2  & 2.8759  & 0.0097  & 3.2422  & 0.0187  \\
    vertex23 & 0.0092  & 2  & 2.8626  & 0.0231  & 3.2446  & 0.0163  \\
    vertex24 & 0.0098  & 5  & 2.8577  & 0.0280  & 3.2207  & 0.0401  \\
    vertex25 & 0.0207  & 3  & 2.9116  & -0.0259  & 3.2178  & 0.0431  \\
    vertex26 & 0.0031  & 3  & 2.9357  & -0.0501  & 3.2195  & 0.0414  \\
    vertex27 & 0.0092  & 2  & 2.7801  & 0.1055  & 3.2367  & 0.0241  \\
    vertex28 & 0.0126  & 4  & 2.7510  & 0.1347  & 3.2264  & 0.0344  \\
    vertex29 & 0.0174  & 3  & 2.7973  & 0.0884  & 3.2371  & 0.0238  \\
    vertex30 & 0.0092  & 4  & 2.7687  & 0.1170  & 3.2242  & 0.0367  \\
    vertex31 & 0.0106  & 4  & 2.7605  & 0.1251  & 3.2361  & 0.0248  \\
    vertex32 & 0.0311  & 6  & 2.8206  & 0.0651  & 3.2225  & 0.0384  \\
    vertex33 & 0.0308  & 12  & 2.6796  & 0.2061  & 3.1929  & 0.0680  \\
    vertex34 & 0.1325  & 17  & 2.6244  & 0.2613  & 3.1893  & 0.0716  \\

    \bottomrule
    \end{tabular}%
  \label{information-loss-karate}%
\end{table}%

The results show that the vertex 33, vertex 34, vertex 1 and vertex 3 are important to the network which is the same as the degree-based structure entropy.
\section{Application}
\label{application}
In this section, the proposed method is used to calculate the structure entropy of the other real networks, namely, the US-airport network \cite{networkdata}, Email networks \cite{networkdata}, the Germany highway networks \cite{nettt}, the US power gird and the protein-protein interaction network in budding yeast \cite{networkdata}. The results are shown in Table \ref{real-networks}.

\begin{table}[H]
  \centering
  \caption{The structure entropy of the real networks}
    \begin{tabular}{llllll}
    \hline
    Network & Nodes  & Edges & $H_{deg}$ & $H_{bet}$ &$H_{partition}$\\
    \hline
    US Airport & 500   & 5962  & 5.025 & 4.7338 &  3.1263 \\
    Email & 1133  & 10902 & 6.631 & 5.5021 & 3.1780\\
    Yeast & 2375  & 23386 & 7.0539 & 6.0931 & 3.0345\\
    US power grid & 4941  & 13188 & 8.3208 & 5.7191 & 1.7018\\
    Germany highway & 1168  & 2486  & 6.9947 & 5.6383  &0.6909\\
    \hline
    \end{tabular}%
  \label{real-networks}%
\end{table}%

The $H_{deg}$ represents the structure entropy which is based on the degree. The $H_{partition}$ represents the structure entropy which is based on the degree partition. The $H_{deg}$ represents the structure entropy which is proposed in the paper.

\begin{figure}[H]
  \centering
  \includegraphics[scale=0.65]{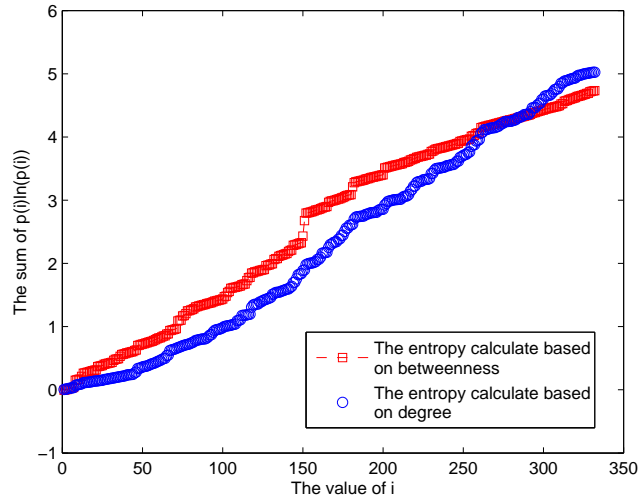}\\
  \caption{The US airport network}\label{USairport}
\end{figure}

\begin{figure}[H]
  \centering
  \includegraphics[scale=0.65]{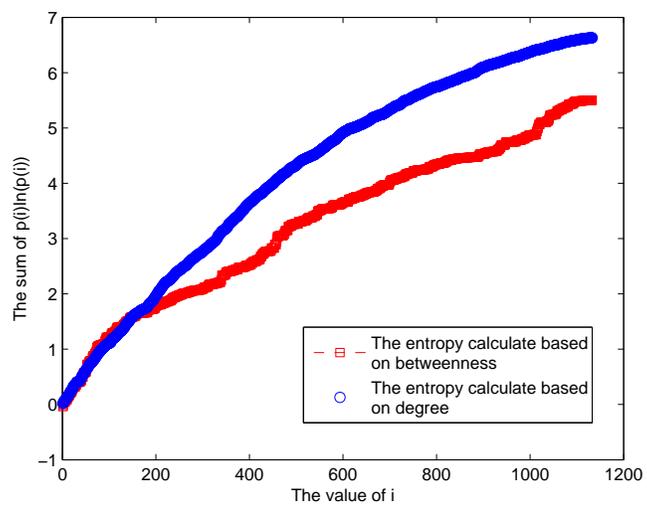}\\
  \caption{The Email network}\label{Email}
\end{figure}

\begin{figure}[H]
  \centering
  \includegraphics[scale=0.65]{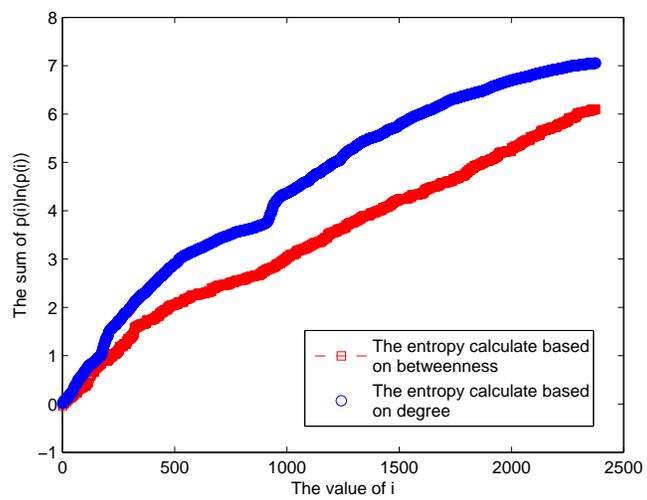}\\
  \caption{The protein-protein interaction network in budding yeast}\label{yeast}
\end{figure}

\begin{figure}[H]
  \centering
  \includegraphics[scale=0.65]{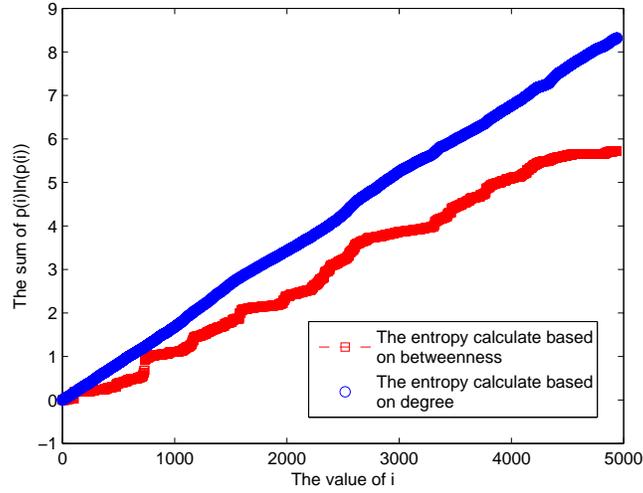}\\
  \caption{The US power grid}\label{power}
\end{figure}

\begin{figure}[H]
  \centering
  \includegraphics[scale=0.65]{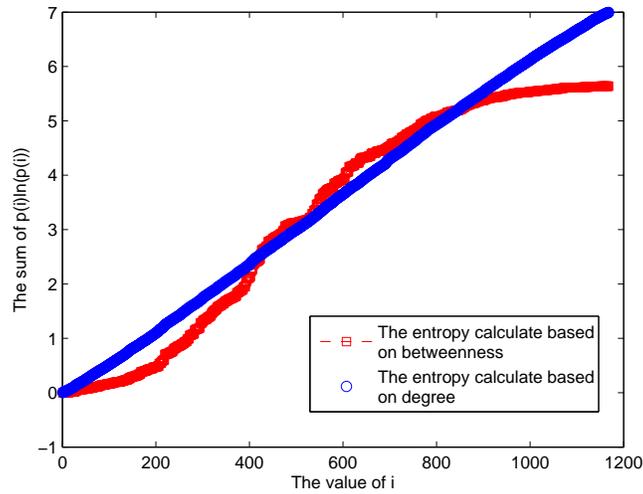}\\
  \caption{The Germany highway network}\label{highway}
\end{figure}

The calculate process of the degree-based structure entropy and the proposed structure entropy are shown in Fig. \ref{USairport}, Fig. \ref{Email}, Fig. \ref{yeast}, Fig. \ref{power} and Fig. \ref{highway}.
\section{Conclusion}
\label{conclusion}
The results of our research reveal that compared with the existing structure entropy the proposed structure entropy is more effective to describe the structure property of the weighted networks.
It is a new method to explore the structure property of the complex networks.
\section*{Acknowledgments}
The work is partially supported by National Natural Science Foundation of China (Grant No. 61174022), Specialized Research Fund for the Doctoral Program of Higher Education (Grant No. 20131102130002), R$\&$D Program of China (2012BAH07B01), National High Technology Research and Development Program of China (863 Program) (Grant No. 2013AA013801), the open funding project of State Key Laboratory of Virtual Reality Technology and Systems, Beihang University (Grant No.BUAA-VR-14KF-02).

%



\bibliographystyle{elsarticle-num}
\bibliography{zqreference}

\begin{thebibliography}{10}
\expandafter\ifx\csname url\endcsname\relax
  \def\url#1{\texttt{#1}}\fi
\expandafter\ifx\csname urlprefix\endcsname\relax\def\urlprefix{URL }\fi
\expandafter\ifx\csname href\endcsname\relax
  \def\href#1#2{#2} \def\path#1{#1}\fi

\bibitem{newman2003structure}
M.~E. Newman, The structure and function of complex networks, SIAM review
  45~(2) (2003) 167--256.

\bibitem{watts1998collective}
D.~J. Watts, S.~H. Strogatz, Collective dynamics of 'small-world' networks,
  nature 393~(6684) (1998) 440--442.

\bibitem{barabasi2000scale}
A.-L. Barab{\'a}si, R.~Albert, H.~Jeong, Scale-free characteristics of random
  networks: the topology of the world-wide web, Physica A: Statistical
  Mechanics and its Applications 281~(1) (2000) 69--77.

\bibitem{berg2002correlated}
J.~Berg, M.~L{\"a}ssig, Correlated random networks, Physical review letters
  89~(22) (2002) 228701.

\bibitem{albert2002statistical}
R.~Albert, A.-L. Barab{\'a}si, Statistical mechanics of complex networks,
  Reviews of modern physics 74~(1) (2002) 47.

\bibitem{barthelemy2004betweenness}
M.~Barthelemy, Betweenness centrality in large complex networks, The European
  Physical Journal B-Condensed Matter and Complex Systems 38~(2) (2004)
  163--168.

\bibitem{shannon2001mathematical}
C.~E. Shannon, A mathematical theory of communication, ACM SIGMOBILE Mobile
  Computing and Communications Review 5~(1) (2001) 3--55.

\bibitem{wang2006entropy}
B.~Wang, H.~Tang, C.~Guo, Z.~Xiu, Entropy optimization of scale-free networks¡¯
  robustness to random failures, Physica A: Statistical Mechanics and its
  Applications 363~(2) (2006) 591--596.

\bibitem{xiao2008symmetry}
Y.-H. Xiao, W.-T. Wu, H.~Wang, M.~Xiong, W.~Wang, Symmetry-based structure
  entropy of complex networks, Physica A: Statistical Mechanics and its
  Applications 387~(11) (2008) 2611--2619.

\bibitem{xu2013degree}
X.-L. Xu, X.-F. Hu, X.-Y. He, Degree dependence entropy descriptor for complex
  networks, Advances in Manufacturing 1~(3) (2013) 284--287.

\bibitem{uci}
Uci network data repository, http://networkdata.ics.uci.edu/data.php?id=105
  (2014).

\bibitem{networkdata}
Pajek datasets, http://vlado.fmf.uni-lj.si/pub/networks/data/ (2014).

\bibitem{nettt}
Tore opsahl, http://toreopsahl.com/datasets/ (2014).

\end{thebibliography}







\end{document}